\documentclass[11pt]{iopart}
\usepackage[english]{babel}
\pdfoutput=1
\usepackage{wasysym}
\usepackage{booktabs}
\usepackage{amssymb}
\usepackage{amsbsy}
\usepackage{verbatim}
\usepackage{graphicx}
\usepackage{epstopdf}
\usepackage{color}
\usepackage{color,soul}
\usepackage{sidecap}
\usepackage{bm}
\usepackage{tikz}

\usepackage[colorlinks,bookmarks=false,citecolor=blue,linkcolor=red,urlcolor=blue]{hyperref}
\usepackage{cancel}

\usepackage{cite}

\catcode`@=11 
\renewcommand\tableofcontents{%
  \section*{\contentsname}%
  \@starttoc{toc}%
}
\catcode`@=12

\def\be{\begin{equation}}
\def\ee{\end{equation}}

\begin{document}

\setlength{\parindent}{0pt}

\title{Negativity spectrum in 1D gapped phases of matter}

\author{Glen Bigan Mbeng, Vincenzo Alba, and Pasquale Calabrese}
\address{$^1$\,SISSA and INFN, via Bonomea 265, 34136 Trieste, Italy. }
\ead{\href{mailto:gmbeng@sissa.it}{gmbeng@sissa.it}, \href{mailto:valba@sissa.it}{valba@sissa.it} 
  and \href{mailto:calabrese@sissa.it}{calabrese@sissa.it}}

\date{\today}


\begin{abstract} 

We investigate the spectrum of the partial transpose (negativity spectrum) of two adjacent regions in gapped one-dimensional models.  
We show that, in the limit of large regions, the negativity spectrum is entirely 
reconstructed from the entanglement spectrum of the bipartite system. 
We exploit this result in the XXZ spin chain, for which the entanglement spectrum is known by means of the corner transfer matrix. 
We find that the negativity spectrum levels are equally spaced, the spacing being half that in the entanglement spectrum. 
Moreover, the degeneracy of the spectrum is described by elegant combinatorial formulas, which are related to the counting 
of integer partitions. 
We also derive the asymptotic distribution of the negativity spectrum. 
We provide exact results for  the logarithmic negativity and for the  moments of the partial transpose. 
They exhibit unusual scaling corrections in the limit $\Delta\to 1^+$ with a corrections exponent which is 
the same as that for the R\'enyi entropies. 

\end{abstract}

\maketitle

\section{Introduction}
\label{intro}

\begin{figure}[t]
\begin{center}
\includegraphics*[width=0.65\linewidth]{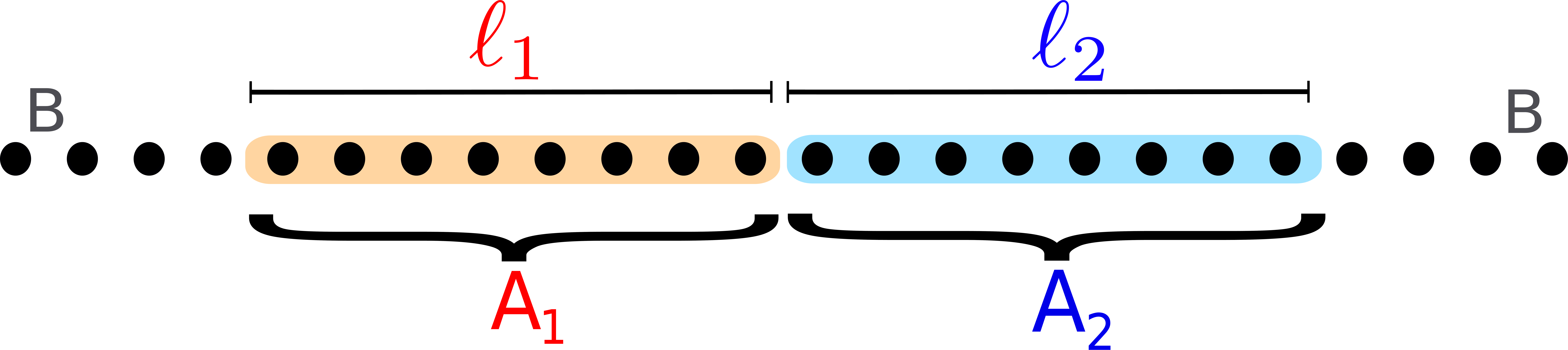}
\end{center}
\caption{ Tripartition of a one-dimensional chain considered in this work. 
 The chain is divided into two adjacent non-complementary intervals 
 $A\equiv A_1\cup A_2$. The rest of the chain is denoted as $B$. Here 
 $\ell_1$ and $\ell_2$ are the lengths of the two intervals. We always 
 consider the situation with $\ell_1=\ell_2=\ell$. 
}
\label{fig0-intro}
\end{figure}

In recent years entanglement witnesses became invaluable diagnostic tools 
to understand complex behaviours in quantum many-body systems~\cite{amico-2008,
calabrese-2009,eisert-2010,laflorencie-2016}. For a bipartite system in a 
pure state, the von Neumann entanglement entropy and the R\'enyi 
entropies are appropriate entanglement measures, but this is not the case for a system in a mixed state. 
For instance, in finite temperature systems 
the entanglement entropy cannot distinguish the quantum correlations 
(entanglement) from the classical ones due to thermal fluctuations. 
A similar issue arises when quantifying the mutual entanglement between 
two non-complementary regions (see Figure~\ref{fig0-intro}). As the two 
intervals are in a mixed state, the von Neumann entropy and the mutual information 
fail to quantify their mutual entanglement. Physically, the mutual information  
cannot distinguish the entanglement between the two intervals 
from the correlations induced by the interaction with the environment.

Although several entanglement measures for mixed states exist, we will focus on the logarithmic negativity~\cite{peres-1996,
zycz-1998,zycz-1999,lee-2000,vidal-2002,plenio-2005} because of its universal features. 
This is defined as follows. Let us consider a system in a pure state $|\psi\rangle$. 
Given a tripartition of the system as $A_1\cup A_2\cup B$, with $A\equiv 
A_1\cup A_2$ the region of interest (as in Figure~\ref{fig0-intro} in one dimension), the reduced 
density matrix $\rho_A$ is defined as $\rho_A\equiv\textrm{Tr}_B\rho$, with $\rho=
|\psi\rangle\langle\psi|$. The logarithmic negativity ${\cal E}_{A_1:A_2}$ is 
\begin{equation}
\label{neg-def}
{\cal E}_{A_1:A_2}\equiv\ln||\rho_{A_1\cup A_2}^{T_2}||_1=\ln\textrm{Tr}|
\rho_{A_1\cup A_2}^{T_2}|. 
\end{equation}
Here $\rho_{A_1\cup A_2}^{T_2}$ is the partially transposed 
reduced density matrix (partial transpose) with respect to $A_2$. This 
is defined from $\rho_A$ as $\langle\varphi_1\varphi_2|\rho^{T_2}_A|\varphi_1'
\varphi_2'\rangle\equiv\langle\varphi_1\varphi_2'|\rho_A|\varphi_1'\varphi_2\rangle$, 
with $\{\varphi_1\}$ and $\{\varphi_2\}$ two bases for $A_1$ and $A_2$, respectively. 
The symbol $||\cdot||_1$ in~\eref{neg-def} denotes the trace norm. In contrast 
with the reduced density matrix, which is positive semidefinite, $\rho_A^{T_2}$ has 
both positive and negative eigenvalues. 
Given the normalisation ${\rm Tr}\rho_{A_1\cup A_2}^{T_2}=1$,   
${\cal E}_{A_1:A_2}$ can be rewritten as a sum over the negative eigenvalues of the partial transpose, 
a property from which the name negativity comes from. 

The logarithmic negativity has been characterised analytically for a generic $1$+$1$ dimensional 
Conformal Field Theory (CFT)~\cite{calabrese-2012,cct-neg-long,calabrese-2013}. 
For two or more disconnected intervals it depends on the full operator content of the 
CFT~\cite{calabrese-2012,cct-neg-long}, similarly to the entanglement entropy~\cite{2int}. 
A remarkable feature is that the negativity is scale invariant at quantum critical points~\cite{hannu-2008,
mrpr-09,hannu-2010,calabrese-2012,cct-neg-long}. Its properties are also known at finite temperature~\cite{calabrese-2015}, 
in CFTs with large central charge~\cite{kpp-14}, disordered spin 
chains~\cite{ruggiero-2016}, out-of-equilibrium models~\cite{ctc-14,hoogeveen-2015,eisler-2014,
wen-2015}, holographic~\cite{rr-15} and  massive quantum field theories~\cite{fournier-2015,bc-16}, 
topologically ordered phases~\cite{lee-2013,castelnovo-2013}, Kondo-like systems~\cite{bayat-2012,
bayat-2014,abab-16}, and Chern-Simons theories~\cite{wen-2016,wen-2016a}. The negativity can 
be calculated analytically for free-bosonic models~\cite{audenaert-2002}, also in $d>1$ 
dimensions~\cite{eisler-2016,dct-16}. However, no results are available for free fermions, 
despite recent progress~\cite{eisler-2014a,coser-2015,ctc-16,chang-2016,hw-16,shapourian-2016,ssr-16,eez-16}. 
The negativity can be obtained numerically in tensor network simulations~\cite{hannu-2008,
calabrese-2013,ruggiero-2016}. Finally, the full spectrum of the partial transpose has been studied 
recently for systems described by a CFT~\cite{ruggiero-2016a} (see Ref.~\citenum{carteret-2005} 
for a proposal to measure the negativity spectrum with quantum circuits). The distribution of 
the positive and negative eigenvalues of the partial transpose has been characterised analytically and,
similar to the eigenvalues of the reduced density matrix~\cite{calabrese-2008} (entanglement spectrum), 
it has been shown that the negativity spectrum is universal and  depends only on the central charge of the CFT. 

%
%

Here we investigate the negativity spectrum in the ground state of one-dimensional gapped 
phases of matter. 
For the entanglement spectrum in gapped phases, a detailed analysis has been already
presented ~\cite{okunishi-1999,alba-2012}. We focus on the 
tripartition with two adjacent intervals embedded in a spin chain (as in Figure~\ref{fig0-intro}). 
Our main result is that for generic gapped systems, in the limit of large intervals, the negativity 
spectrum can be constructed entirely from the entanglement spectrum of the bipartition of the system in two semi-infinite halves. 
This reflects two important physical facts. 
First, the negativity spectrum is determined by degrees of freedom living at the boundaries between the subsystems,
similarly to what happens for the entanglement spectrum~\cite{alba-2012}. 
Second, due to the finite bulk correlation 
length, near the boundary between $A_1$ and $A_2$ the subsystem $A_1\cup A_2$ can be effectively described by a pure state. 

We exploit these ideas to derive the negativity spectrum of the XXZ spin chain from its 
entanglement spectrum, which is known exactly \cite{okunishi-1999,alba-2012,n-95,pkl-99,cc-04,rp-09} 
from Corner Transfer Matrix (CTM) techniques. 
The anisotropic Heisenberg spin chain (XXZ chain) is defined by the Hamiltonian 
\begin{equation}
\label{ham}
H_{XXZ}=\sum\limits_{i=1}^L[S_i^xS_{i+1}^x+S_i^yS_{i+1}^y+\Delta 
S_i^zS_{i+1}^z],
\end{equation}
where $S^\alpha_i\equiv\sigma^{\alpha}_i/2$, with $\sigma^\alpha_i$ the Pauli matrices, are 
spin-$\frac{1}{2}$ operators acting on the site $i$ of the chain, and $\Delta$ the anisotropy parameter. 
For $-1<\Delta\le 1$ the spectrum of the XXZ chain is gapless, and its low-energy part is described by a $c=1$ CFT. 
For $\Delta>1$, which is the regime of interest here, the model is in the gapped antiferromagnetic regime.

The manuscript is organised as follows. In section~\ref{neg_gapped} we show the construction of the partial 
transpose in the ground state of generic gapped systems. Section~\ref{moments} describes how the negativity 
spectrum is constructed starting from the moments of the partial transpose and the entanglement spectrum of 
a bipartition. In section~\ref{ns-xxz} we use these results to derive the structure of the negativity 
spectrum of the gapped XXZ chain. In section~\ref{deg-patt} we present some analytic results for the 
logarithmic negativity and the moments of the partial transpose. 
In section~\ref{unusual-sec} we analyse the scaling of the negativity and of the moments of the partial transpose  in the limit $\Delta\to1^+$, 
and investigate the presence of unusual corrections to the scaling. In section~\ref{neg-dist} we derive analytic results 
for the distribution of the negativity spectrum in the XXZ chain. Our conclusions are in section~\ref{concl}.

\section{Partially transposed reduced density matrix in gapped phases} 
\label{neg_gapped}

%
\begin{figure}[t]
\begin{center}
\includegraphics*[width=0.65\linewidth]{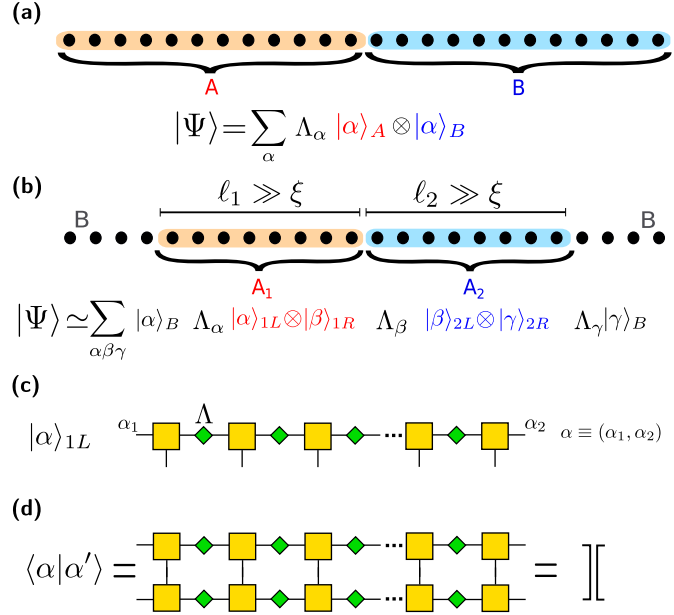}
\end{center}
\caption{ Logarithmic negativity of two adjacent intervals $A_1$ and $A_2$. 
 (a) A bipartition of the chain as $A\cup B$ with the Schmidt decomposition of the ground-state wavefunction. 
 $|\alpha\rangle_{A(B)}$ are two orthonormal bases for part $A(B)$, $\Lambda_\alpha$ are the Schmidt values. 
 (b) The tripartition used in this work. The length of the two intervals $\ell_1(\ell_2)$ is much larger than the correlation 
 length $\xi$. 
 In this regime, the entanglement is ``localised'' close to the three edges and  an approximate decomposition for this tripartition 
 is reported below the cartoon. 
 Here the $\Lambda$s are the Schmidt coefficients for the three bipartitions in semi-infinite chains at each edge (alike in (a)),
 while $|\alpha\rangle_{1L}$ and $|\beta\rangle_{1R}$ are the two Schmidt bases for the 
 Hilbert space around the left and right boundary of $A_1$ (and similarly the others kets). 
 (c) Matrix Product State representation for $|\alpha\rangle_{1L}$. The squares denote the matrices $T$ (cf.~\eref{mps}), 
 rhombi the matrix $\Lambda$. (d) Orthogonality of the basis $|\alpha\rangle_{1L}$. For $\ell_1
 \gg\xi$, one has $\langle\alpha|\alpha'\rangle=\delta_{\alpha\alpha'}$. 
}
\label{fig0}
\end{figure}
%
In this section we show that in the limit of large subsystems the negativity spectrum of two adjacent 
intervals can be reconstructed entirely from the entanglement spectrum of the bipartition of the system in two semi-infinite halves. 
The main physical idea is very simple and a very similar one was used for the negativity in \cite{fournier-2015}. 
Since we are interested in a tripartition in the regime in which both the finite intervals $\ell_{1,2}$ are much larger than 
the finite correlation length $\xi$, the entanglement is localised  around the three edges of the tripartition, i.e. 
at the boundary between $B$ and $A_1$, that between $A_1$ and  $A_2$, and between $A_2$ and $B$ (see Fig. \ref{fig0}).
Consequently, the state is well approximated (with exponential accuracy in $\ell_{i}/\xi$) by a triple sum on the three bipartitions
(which involves the same Schmidt coefficients)
and the negativity spectrum is simply written in terms of the entanglement spectrum. 
Having explained the main idea, in the following we make the argument more rigorous employing  
the language of Matrix Product States (MPS). We restrict ourselves to spin-$\frac{1}{2}$ models 
(spin chains) with open boundary conditions, but our arguments are straightforwardly generalised to other situations.

%
\begin{figure}[t]
\begin{center}
\includegraphics*[width=0.75\linewidth]{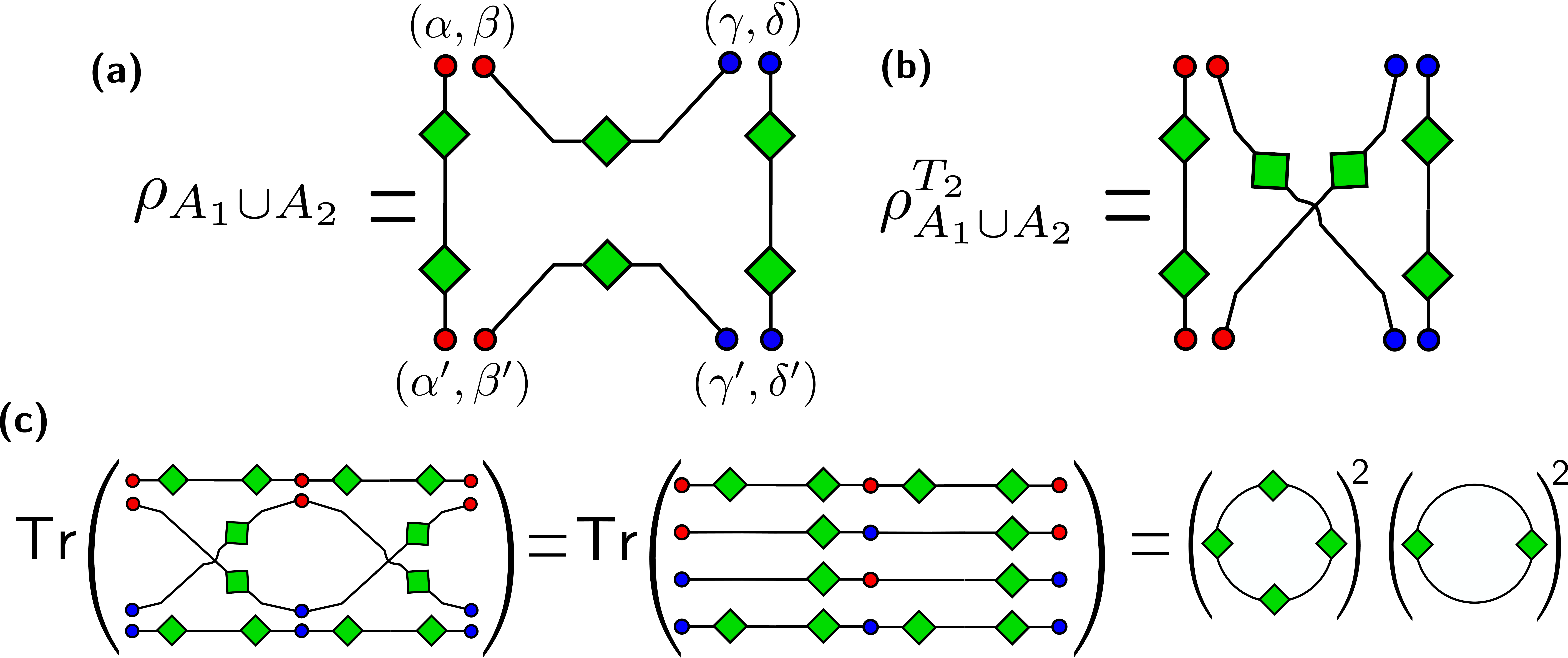}
\end{center}
\caption{ Matrix Product State representation for the reduced density matrix 
 $\rho_{A_1\cup A_2}$ and the partial transpose $\rho_{A_1\cup A_2}^{T_2}$. 
 The rhombi are the diagonal matrices containing the singular values 
 (entanglement spectrum). Note in (b) that the effect of the 
 partial transposition is to swap $\gamma\leftrightarrow\gamma'$. The partial 
 transposition is ineffective on $\delta\delta'$. (c) The second moment 
 of the partial transpose: pictorial proof of the identity 
 $\textrm{Tr}(\rho_{A_1\cup A_2}^{T_2})^2=(\textrm{Tr}\rho^2_{A_1})^2
 \textrm{Tr}(\rho_{A_1})^2$. Note the untwisting of the tensor lines 
 in the intermediate step. 
}
\label{fig1}
\end{figure}
%

In the MPS framework the ground state of a generic spin chain is represented as 
\begin{equation}
\label{mps}
|\Psi\rangle=\sum\limits_{\sigma_1,\dots,\sigma_L}T_{\nu_1}^{\sigma_1}T_{\nu_1,\nu_2}^{\sigma_2}\cdots 
T_{\nu_L}^{\sigma_L}|\sigma_1,\dots,\sigma_L\rangle. 
\end{equation}
Here $\sigma_i=\pm 1$ are the  physical indices, which represent the states in the local 
Hilbert space, while $1\le\nu_i\le\chi_i$ are the virtual indices, with $\chi_i$ the local bond 
dimension of the MPS. In~\eref{mps}, $T_{\nu_1}^{\sigma_1}$ and $T_{\nu_L}^{\sigma_L}$ are vectors, 
whereas for $1<i<L$, $T_{\nu_i,\nu_{i+1}}^{\sigma_i}$ are $\chi_i\times\chi_i$ matrices. 
The sum over repeated indices is assumed. Given a  bipartition of the chain as $A\cup B$ 
(see Figure~\ref{fig0} (a)),  the  Schmidt decomposition allows us to rewrite $|\Psi\rangle$ as 
\begin{equation}
\label{sch_bip}
|\Psi\rangle=\sum\limits_\alpha \Lambda_\alpha|\alpha\rangle_A\otimes|\alpha\rangle_B. 
\end{equation}
Here $|\alpha\rangle_{A(B)}$ are orthonormal bases (Schmidt vectors) for parts $A$ and $B$ and
$\Lambda_\alpha$ is the diagonal matrix with the Schmidt values. The 
non-zero eigenvalues of the reduced density matrix $\rho_A$ are given as $\Lambda_\alpha^2$. 
For translational invariant systems (for instance for the half-infinite chain), 
the tensors $T_{\nu_i,\nu_{i+1}}^{\sigma_{i}}$ do not depend on the site $i$. 

Let us now consider the tripartition of the chain $A_1\cup A_2\cup B$  as in Figure~\ref{fig0} (b), 
i.e., two adjacent intervals and the rest, with the lengths of the two intervals being much 
larger than the bulk correlation length $\xi$. 
The idea of the approach is to construct an approximate basis for the tripartite system. 
For gapped systems the eigenvectors of the reduced density matrix and of its
partial transpose are expected to be localised at the boundary between the two 
subsystems and between the subsystems and the rest. 
This suggests that a useful 
orthonormal basis for $A_1$ can be constructed as $|\alpha\rangle_{1L}\otimes|\beta
\rangle_{1R}$, with $|\alpha\rangle_{1L}$ and $|\beta\rangle_{1R}$ the Schmidt vectors 
appearing in~\eref{sch_bip} (see also Figure~\ref{fig0}). The tensor product structure of 
the basis reflects the factorisation of the two edges of $A_1$, which 
relies on the condition $\ell_1\gg\xi$. Similarly, for $A_2$ one can construct 
the orthonormal basis as $|\beta\rangle_{2L}\otimes|\gamma\rangle_{2R}$, where, again, 
$|\beta\rangle_{2L}$ and $|\gamma\rangle_{2R}$ are the same Schmidt vectors in~\eref{sch_bip}. 
This allows one to decompose the wavefunction of the system as 
\begin{equation}
\label{sch_trip}
|\Psi\rangle\simeq\sum\limits_{\alpha,\beta,\gamma}|\alpha\rangle_B\Lambda_\alpha|\alpha\rangle_{1L}
\otimes|\beta\rangle_{1R}\Lambda_\beta|\beta\rangle_{2L}\otimes|\gamma\rangle_{2R}\Lambda_\gamma 
|\gamma\rangle_{B}, 
\end{equation}
where $|\alpha\rangle_B$ and $|\gamma\rangle_B$ are elements of the Schmidt basis for part $B$ 
of the chain and $\Lambda_\alpha$ is the Schmidt coefficient of the bipartition~\eref{sch_bip}. 
Eq.~\eref{sch_trip}, in the regime $\ell_1,\ell_2\gg\xi$, has an exponential accuracy, i.e. the error is as small as $e^{-\ell_i/\xi}$, 
as  argued for the negativity in Ref. \citenum{fournier-2015}.

The decomposition~\eref{sch_trip} of $|\Psi\rangle$ is illustrated pictorially in Figure~\ref{fig0} (c), 
using the MPS formalism to represent the vectors $|\alpha\rangle_{1L}$. 
Similar representations hold for the other vectors appearing in~\eref{sch_trip}. 
The square boxes in Figure~\ref{fig0} (c) represent the tensors $T_{\nu_i,\nu_{i+1}}^{\sigma_{i+1}}$ (cf.~\eref{mps}), 
the rhombi the diagonal matrix $\Lambda_\alpha$ as in~\eref{sch_bip}. The vertical dangling bonds are the 
physical indices $\sigma_i$ and are summed over. The index $\alpha$ is a composite index $\alpha\equiv
(\alpha_1,\alpha_2)$, with $\alpha_1,\alpha_2$  the virtual indices at the edges of the MPS. The length 
of the MPS (number of tensor in Figure~\ref{fig0} (c)) has to be much larger than the correlation length.  
Note that since $|\alpha\rangle_{1L}$ are constructed from the Schmidt vectors for the bipartition, they 
are (approximately) orthonormal. The orthonormality condition $\langle\alpha|\alpha'\rangle=\delta_{\alpha,
\alpha'}$ is illustrated in Figure~\ref{fig0} (d). 

Using~\eref{sch_trip}, it is straightforward to derive the MPS representation of the reduced density matrix 
$\rho_{A_1\cup A_2}$. This is reported in Figure~\ref{fig1} (a). The composite indices $(\alpha,\beta)$ and 
$(\alpha',\beta')$ refer to $A_1$, whereas $(\gamma,\delta)$ $(\gamma',\delta')$ are for $A_2$. 
Specifically, the indices $\beta,\gamma$ refer to the degrees of freedom living at the 
interface between $A_1$ and $A_2$, while $\alpha,\delta$ to the ones lying at the interface between $A_1\cup A_2$ 
and the rest of the system. Again, the rhombi denote the Schmidt values $\Lambda_{\alpha}$ 
for the bipartite system (Figure~\ref{fig0} (a)). Clearly, $\rho_{A_1\cup 
A_2}$ is a tensor product of three terms, corresponding to the three different boundaries present in the 
system. At this point the partial transpose with respect to $A_2$ is trivially derived since it acts 
as a global transposition on the indices $\alpha,\alpha'$ and $\delta,\delta'$. 
The only non trivial effect is the swap $\gamma\leftrightarrow\gamma'$. 
The structure of $\rho_{A_1\cup A_2}^{T_2}$ is illustrated pictorially in Figure~\ref{fig1} (b).

\section{Negativity spectrum from the moments of the partial transpose} 
\label{moments}

In the above section we showed that the partial transpose of two adjacent intervals (in a tripartite system) can be 
written in terms of the reduced density matrix of a semi-infinite half of the gapped system.
Thus a similar relation should holds also between the entire negativity spectrum of the tripartition 
and the entanglement spectrum of the bipartition.
This relation is obtained in an easy way exploiting the  moments $\textrm{Tr}(\rho_{A_1\cup A_2}^{T_2})^n$ 
that can be also written in terms of the moments of the reduced density matrix 
for the bipartition, as in Figure~\ref{fig0} (a). 
By very simple algebra, it is straightforward to derive the following identities depending on the parity of $n$:
\begin{equation}
\label{constr}
\textrm{Tr}(\rho_{A_1\cup A_2}^{T_2})^{n}=
(\textrm{Tr}\rho^n_{A_1})^2\times
\left\{
\begin{array}{cc}
(\textrm{Tr}\rho^{n_e/2}_{A_1})^2 & n_e\;\;\textrm{even},\\
\textrm{Tr}\rho^{n_o}_{A_1}       & n_o\;\;\textrm{odd}, 
\end{array}
\right.
\end{equation}
where $n_e$ and $n_o$ are even and odd integers, respectively. A pictorial derivation of~\eref{constr} 
for $n=2$ is reported in Figure~\ref{fig1} (c). The contributions of the straight horizontal lines 
 is $(\textrm{Tr}(\rho_{A_1})^2)^2$ (top and bottom lines in the left member in Figure~\ref{fig1} (c)). 
This is clearly generalised as $(\textrm{Tr}\rho^n_{A_1})^2$ for arbitrary $n$, and it does not 
depend on the parity of $n$. The non-trivial contribution to the moments originates from the 
``entangled'' lines in Figure~\ref{fig1} (c). For $n=2$, in taking the trace the two lines can be 
``disentangled'' into two independent circles (second term in the rightmost member in the Figure). 
For generic even $n$ one obtains two independent closed diagrams (``circles''), each containing $n$ 
rhombi. This corresponds to $(\textrm{Tr}\rho_{A_1}^{n/2})^2$. On the other hand, for $n$ odd 
the ``entangled'' diagrams are topologically equivalent to a single closed circle  with $2n$ 
rhombi, which is $\textrm{Tr}\rho_{A_1}^n$. 
The equations in~\eref{constr} are similar to the relations valid for the moments of the partial transpose for a bipartite {\it pure}
state  \cite{calabrese-2012,cct-neg-long}
\begin{equation}
\label{constr-pure}
\textrm{Tr}(\rho_{A_1\cup A_2}^{T_2})^{n}=
\left\{
\begin{array}{lc}
(\textrm{Tr}\rho^{n_e/2}_{A_1})^2 & n_e\;\textrm{even},\\
\textrm{Tr}\rho^{n_o}_{A_1}       & n_o\;\textrm{odd}. 
\end{array}
\right.
\end{equation}
Clearly, the relations~\eref{constr-pure} are the same as~\eref{constr} apart for a multiplicative 
factor. This is physically expected because for $\ell_1,\ell_2\gg\xi$ the system around the boundary 
between $A_1$ and $A_2$ is effectively in a pure state. Moreover, Eq.~\eref{constr} shows
that $\textrm{Tr}(\rho_{A_1\cup A_2}^{T_2})^n$ depends on the degrees of freedom at the 
boundary between $A$ and $B$ through the term $(\textrm{Tr}\rho_{A_1}^n)^2$. 
This reflects that for $n\neq1$ the moments of the partial transpose are not good measures of 
the {\it mutual} entanglement between $A_1$ and $A_2$. Note, however, that the logarithmic 
negativity, which is obtained from~\eref{constr} by taking the limit $n_e\to 1$, depends only on 
the degrees of freedom at the boundary between $A_1$ and $A_2$, as it should.  

Finally, the relations in~\eref{constr} imply that the negativity spectrum can be completely 
derived from the entanglement spectrum of a bipartite system, since they represent an infinite set of 
equations for the eigenvalues of the partial transpose. 
Denoting as $\mu_i$ the eigenvalues of the  reduced density matrix $\rho_{A_1}$ of the half-infinite system, 
the eigenvalues of the partial transpose $\lambda_{i,j,k,l}$ can be written as 
\begin{equation}
\label{neg_spec}
\lambda_{i,j,k,l}=\mu_i\mu_j\left\{
\begin{array}{ll}
\sqrt{\mu_k\mu_l} & k<l,\\
\mu_k             & k=l,\\
-\sqrt{\mu_k\mu_l} & k>l. 
\end{array}
\right.
\end{equation}
This can be proved by verifying that the $\lambda_{i,j,k,l}$ defined in~\eref{neg_spec} satisfy~\eref{constr}. 

\section{Structure of the negativity spectrum in the gapped Heisenberg chain} 
\label{ns-xxz}

The goal of this section is to apply~\eref{neg_spec} to derive the negativity spectrum of the gapped 
antiferromagnetic XXZ chain. To this purpose, the key ingredient is the half-chain entanglement spectrum, 
which can be constructed completely using the Corner Transfer Matrix (CTM) techniques~\cite{baxter-book}.

\subsection{Entanglement spectrum of the XXZ chain via the Corner Transfer Matrix}
\label{es}

In this section we review the construction of the entanglement spectrum using the CTM approach 
\cite{pkl-99,cc-04,rp-09}. Given a two dimensional statistical physics model, the CTM $\hat M$ 
is defined as the transfer matrix connecting a horizontal row to a vertical one. Specifically, 
the matrix elements of $\hat M$ are the partition functions on the lower left quadrant with 
fixed boundary conditions on the $x$ and $y$ axis. The reduced density matrix $\rho_A$ of 
the half-infinite system is related to $\hat M$ as 
\begin{equation}
\rho_A=\frac{\hat M^4}{\textrm{Tr}\hat M^4}. 
\end{equation}
For the gapped XXZ chain, in the infinite lattice limit, the reduced density matrix can be 
rewritten as 
\begin{equation}
\rho_A=\frac{e^{-H_{CTM}}}{\textrm{Tr}e^{-H_{CTM}}},
\end{equation}
where $H_{CTM}$ is a modular (or entanglement) Hamiltonian. $H_{CTM}$ can be recast in the free-fermion 
form  
\begin{equation}
\label{hctm}
H_{CTM}= 
\sum_{j=0}^\infty\epsilon_j \hat n_j, 
\end{equation}
with  
$\hat n_j$ a fermionic number operator with eigenvalues $0$ and $1$, 
and $\epsilon_j$ the ``single-particle''  entanglement spectrum levels \cite{pkl-99}
\begin{equation}
\label{spes}
\epsilon_j= 2j\epsilon,\quad j\in\mathbb{N},\quad\textrm{and}\;\; \epsilon\equiv 
\textrm{arccosh} \Delta. 
\end{equation}
The entanglement spectrum is obtained by filling in all the possible ways the single particle levels $\epsilon_j$ 
in~\eref{hctm} (i.e., setting all $\hat n_j$ equal either to $0$ or $1$). 
The resulting levels are equally spaced with spacing $2\epsilon$. 
It is also clear from~\eref{hctm} and~\eref{spes} that these levels are highly degenerate. 
The degeneracy of the level $\zeta=2\epsilon n$ with $n=\sum_j j$ (cf.~\eref{spes}) is obtained by counting the number of 
ways of obtaining $n$ as a sum of smaller non-repeated integers (including zero). 
This is the problem of counting the number $q(n)$ of (restricted) partitions of $n$. 
The number $q(n)$ is conveniently generated as a function of $n$ via the generating function
\begin{equation}
\label{g}
G(z)\equiv\sum_{k=0}^{\infty}q(k)z^k=\prod_{k=1}^{\infty}\left(1+z^k \right). 
\end{equation}
Thus, $q(n)$ is obtained from~\eref{g} as the coefficient of the monomial $z^n$. 
The ES degeneracy is given by $2q(n)$, with the factor two coming from the overall double degeneracy given by 
the zero mode $n_0=0,1$ with $\epsilon_0=0$.
\footnote{Physically this is due to the fact that in our construction the CTM selects the (equally probable) linear combination of the two
degenerate ground states of the XXZ spin chain for $\Delta>1$ (i.e. Neel plus AntiNeel at $\Delta=\infty$). 
The degeneracy of the ES levels for symmetry breaking states is just $q(n)$ as in Ref. \citenum{alba-2012}. 
In the CTM approach, the symmetry breaking states can be obtained from \eref{hctm} by letting the sum starting for $j=1$ instead of 
$j=0$, which correspond to fixing a boundary condition at the origin \cite{baxter-book}.
}  
Clearly, the ES degeneracy is given by $2q(n)$ for those bipartitions which have one boundary between 
$A$ and $B$ (two half-infinite lines or finite ones with lengths $\ell_i\gg \xi$). 
For bipartitions with two boundaries (e.g. an interval embedded in an infinite or a finite system) the degeneracies  
are more complicated~\cite{alba-2012}. 

We mention that the corner transfer matrix techniques have been applied to the study of entanglement entropy and 
spectrum also to other exactly solvable models \cite{w-06,eer-10,eefr-12,df-13,br-16}.

%
\begin{figure}[t]
\begin{center}
\includegraphics*[width=0.65\linewidth]{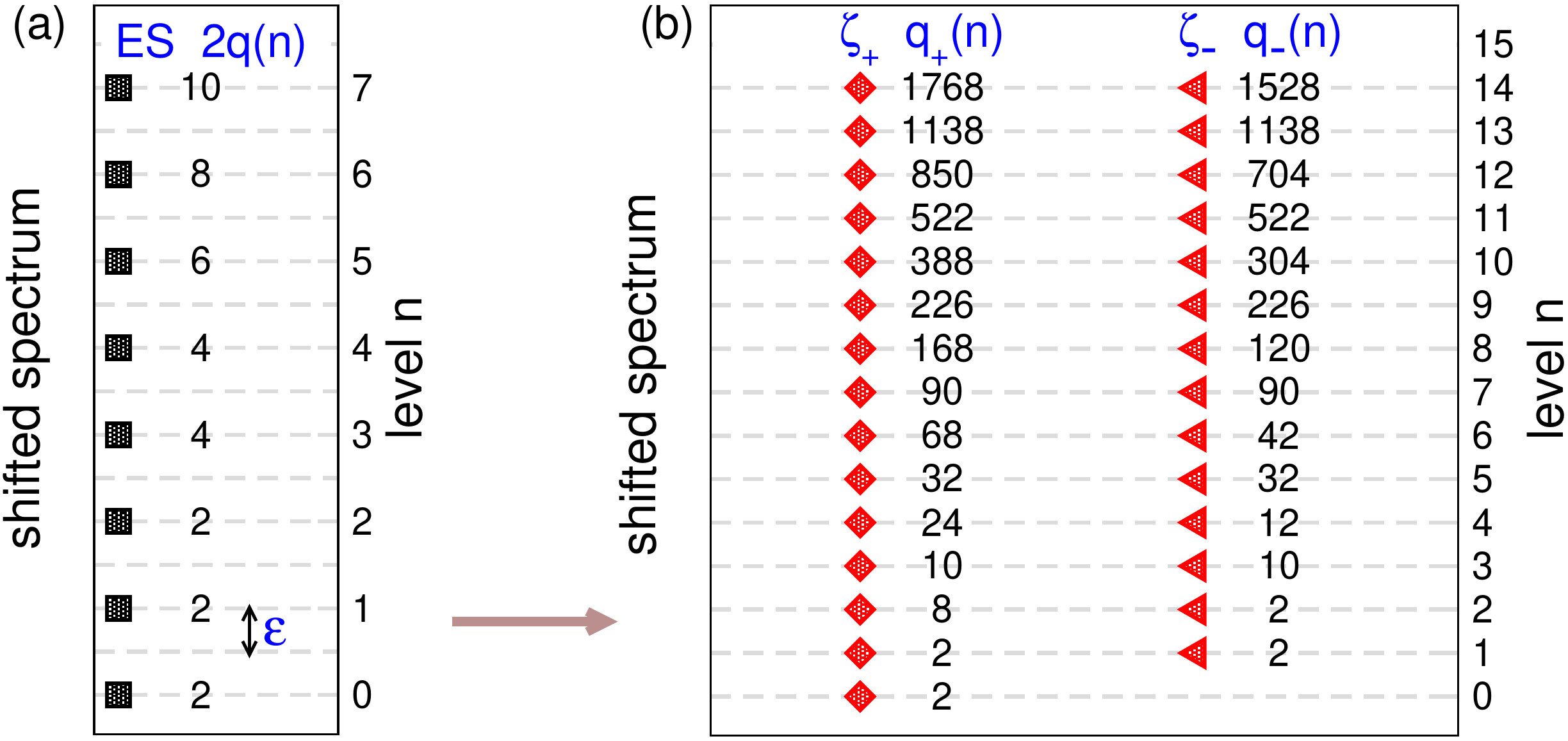}
\end{center}
\caption{Survey of theoretical results for the negativity spectrum of two adjacent intervals in the gapped XXZ spin chain. 
 The cartoon shows the levels $\zeta_\pm$ in the positive and negative sectors of the 
 spectrum (rhombi and triangles, respectively). 
 The entanglement spectrum levels for two half-infinite lines are reported for comparison (squares). 
 All the spectra are shifted by the lowest level. 
 The numbers accompanying the  symbols denote the degeneracy. 
 For the ES this is $2q(n)$, with $q(n)$ the number of distinct integer partitions of $n$. 
 $q_\pm$ denotes the degeneracy of $\zeta_\pm$, with $q_+(n)=q_-(n)$ for odd $n$. 
 Both for the negativity and the entanglement spectrum, the levels are equally spaced, the spacing being 
 $\epsilon\equiv\textrm{arccosh} \Delta$ and $2\epsilon$, respectively. 
}
\label{fig5}
\end{figure}
%

\subsection{Negativity spectrum}
\label{nes}

Here we derive the structure of the negativity spectrum in the gapped Heisenberg spin 
chain for two adjacent finite intervals of lengths much larger than the correlation length.
We denote by $\zeta_+=-\ln \lambda_+$ all the levels corresponding to the positive eigenvalues $\lambda_+$ of the 
partial transpose and by $\zeta_-=-\ln |\lambda_-|$ those corresponding to the negative eigenvalues $\lambda_-$. 
Eq.~\eref{neg_spec} implies that both $\zeta_+$ and $\zeta_-$ are equally spaced, the spacing 
being $\epsilon$ (cf.~\eref{spes}), while for the entanglement spectrum the spacing  is $2\epsilon$. 
This difference is due to the square root in~\eref{neg_spec}. 

The structure of the negativity spectrum is presented in Figure~\ref{fig5}. The Figure   
shows on the left the entanglement spectrum levels (square symbols), shifted by the lowest level. 
The accompanying numbers are the level 
degeneracies, which are given as $2q(n)$, with $q(n)$ the number of restricted integer 
partitions of $n$. 

The rhombi and the triangles show the negativity spectrum levels corresponding to 
positive and negative eigenvalues of $\rho_{A_1\cup A_2}^{T_2}$.  
The corresponding degeneracies $q_\pm(n)$ are also reported. 
Both $\zeta_+$ and $\zeta_-$ are shifted by the lowest level. 
The degeneracy sequences $q_\pm(n)$ are constructed from~\eref{neg_spec}, and they 
are conveniently derived from the generating functions 
\begin{equation}
\label{gf}
G_\pm(x)\equiv\sum_{k=0}^\infty q_\pm(k)x^k=G^2(x^2)[G^2(x)\pm G(x^2)]. 
\end{equation}
Here $G(x)$ is the same generating function as in~\eref{g}. The terms $G^2(x^2)$ and 
$G^2(x)$ in~\eref{gf} correspond to $\mu_i\mu_j$ and $\sqrt{\mu_k\mu_l}$ in~\eref{neg_spec}, 
respectively. The second term in the square bracket in~\eref{gf} properly counts the 
diagonal term with $k=l$ in~\eref{neg_spec}. A non-trivial consequence of~\eref{gf} is that 
$q_+(n)=q_-(n)$ for $n$ odd, as obvious from Figure~\ref{fig5}. 

Thus the negativity spectrum levels can be written as
\begin{equation}
\label{pred}
\zeta=\zeta_0+n\epsilon,\quad n=0,1,\dots,
\end{equation}
with $\zeta_0$ the lowest levels which is obtained by imposing
the normalisation $\textrm{Tr}\rho^{T_2}_{A_1\cup A_2}=1$. Indeed we have 
\begin{equation}
\textrm{Tr}\rho^{T_2}_{A_1\cup A_2}= \sum \lambda_++\sum \lambda_-=
e^{-\zeta_0}\sum_{k=0}^\infty[q_+(k)-q_-(k)]e^{-k\epsilon}=1, 
\label{norm}
\end{equation}
which implies 
\begin{equation}
\label{zeta0}
\fl \zeta_0=\ln\sum_{k=0}^\infty[q_+(k)-q_-(k)]e^{-k\epsilon}=
\ln \left[G_{+}(e^{-\epsilon})- G_{-}(e^{-\epsilon}) \right] =3\ln(G(e^{-2\epsilon}))+\ln 2, 
\end{equation}
with $G_\pm(z)$ defined in~\eref{gf}. It follows then that the largest eigenvalue of $\textrm{Tr}\rho^{T_2}_{A_1\cup A_2}$ is 
\begin{equation}
\lambda_0=\frac1{2(G(e^{-2\epsilon}))^3}\,.
\label{lambda0}
\end{equation}

\subsection{Numerical checks using TEBD} 
\label{num-checks}

%
\begin{figure}[t]
\begin{center}
\includegraphics*[width=0.65\linewidth]{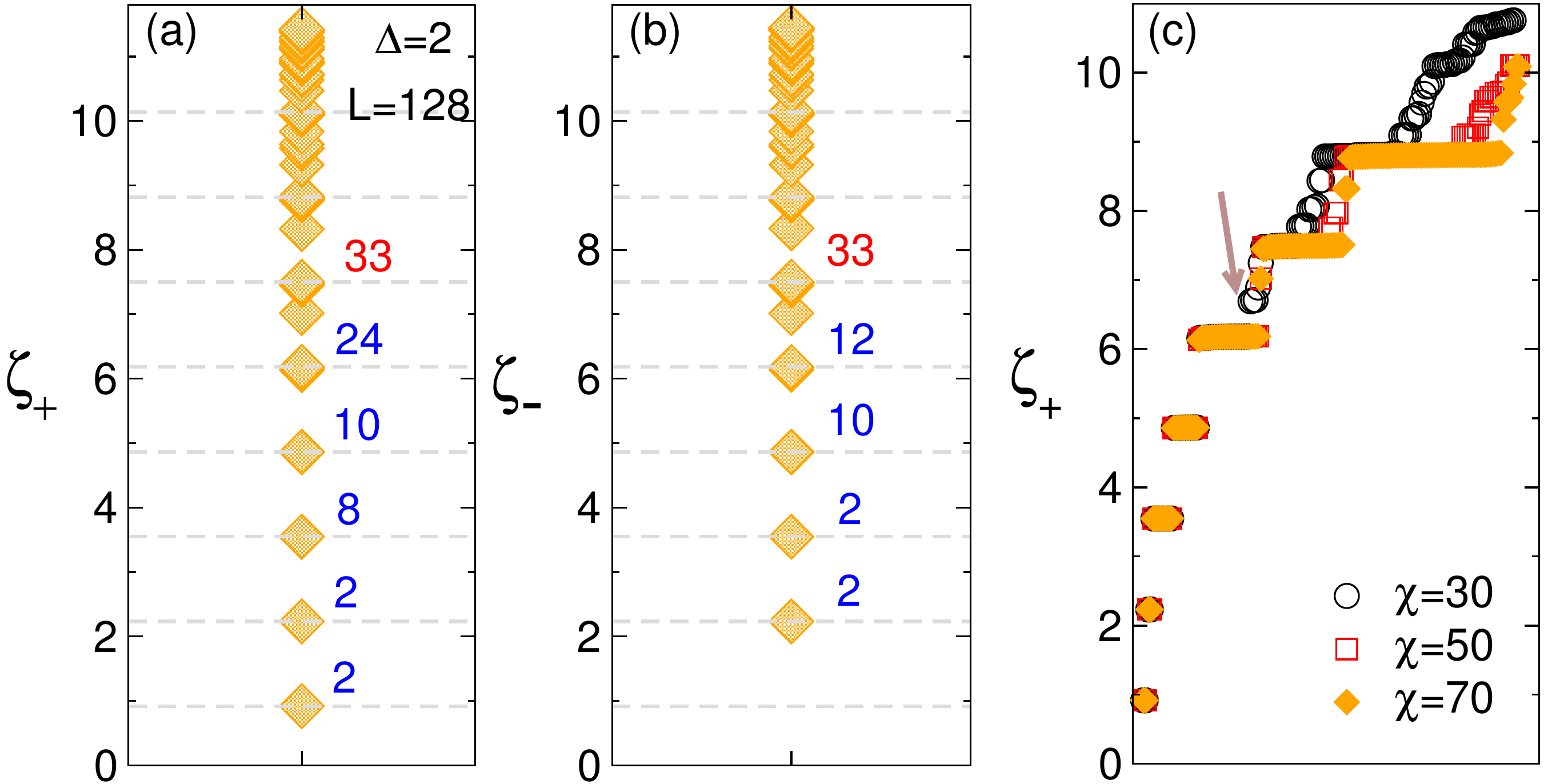}
\end{center}
\caption{TEBD results for the negativity spectrum of two adjacent intervals in the gapped XXZ spin chain. 
 The data are for a chain with $L=128$, the intervals $\ell_1=\ell_2=L/4$, and $\Delta=2$. 
 (a)(b) Negativity spectrum levels $\zeta_\pm$ in the sector of positive and negative eigenvalues 
 of $\rho_A^{T_2}$. The accompanying numbers are the level 
 degeneracies. The horizontal dashed lines are the analytical 
 results for large intervals. (c)  $\zeta_+$ 
 at $\Delta=2$ and $L=128$, obtained using different values of 
 the bond dimension $\chi=30,50,70$. Finite $\chi$ effects are 
 visible in the higher part of the spectrum (arrow). 
}
\label{fig6}
\end{figure}

We now provide some numerical checks of the results presented for both the values and the degeneracy of the 
negativity spectrum level summarised in Figure~\ref{fig5}. 
Figure~\ref{fig6} shows  time evolving block decimation~\cite{vidal1,vidal2} (TEBD) data for the negativity spectrum of the XXZ chain at $\Delta=2$. 
The data are for a chain with $L=128$ sites and open boundary conditions. 
The two blocks $A_1$ and $A_2$ are both composed of  $L/4$ spins. 
The simulations in Figure~\ref{fig6} (a) and (b) are performed at fixed bond dimension $\chi=70$. 
The ground state of the $XXZ$ chain is obtained by imaginary time evolution of an initial MPS via 
the Trotter decomposition of, as usual in TEBD. 
For a generic tripartition, the partial transpose and the negativity spectrum are constructed using the method 
described in Ref.~\citenum{ruggiero-2016}, which relies on the MPS representation of the 
ground state of the XXZ chain. 
The calculation of the negativity spectrum involves the diagonalisation of a $\chi^2\times\chi^2$ matrix. 
The computational cost is $\chi^6$. In our setup, to avoid boundary effects, the two intervals $A_1\cup A_2$ 
are placed at the centre of the chain. 
Figures~\ref{fig6} (a) and (b) show the positive and negative levels $\zeta_+$ and $\zeta_-$ of the negativity spectrum. 
The horizontal dash-dotted line are the expected results in the thermodynamic limit (\ref{pred})
and they match the numerical data perfectly in the lower part of the spectrum. 
The accompanying numbers are the spectrum degeneracies as obtained from TEBD. 
For the lower part of the negativity spectrum ($\zeta\lesssim 6$), the degeneracy counting 
is in agreement with the expected result in the thermodynamic limit reported in Figure~\ref{fig5}. For 
$\zeta\gtrsim 6$ deviations are present. This is due to both the finite bond dimension of the TEBD and the 
finite size of the chain. Similar finite-size effects are observed in the entanglement spectrum of the 
XXZ chain~\cite{alba-2012}. Finite $\chi$ effects are illustrated in Figure~\ref{fig6} (c) focusing 
on the positive part of the negativity spectrum. The different symbols are TEBD results for different 
values of $\chi$. The negativity spectrum levels clearly converged for $\zeta\lesssim 6$, while 
higher levels vary upon changing $\chi$.

\section{Moments of the partial transpose and logarithmic negativity} 
\label{deg-patt}

The knowledge of the negativity spectrum can be used to provide exact results for the 
moments of the partial transpose and the logarithmic negativity in the gapped XXZ chain, 
as we are going to do now. 
By definition, the logarithmic negativity of two adjacent intervals is 
\begin{equation}
\label{neg}\fl 
{\cal E}_{A_1:A_2}\equiv
\textrm{Tr}|\rho^{T_2}_{A_1\cup A_2}|= \sum \lambda_+- \sum \lambda_-= e^{-\zeta_0}\sum_{k=0}^\infty[q_+(k)+
q_-(k)]e^{-k\epsilon}.
\end{equation}
The sum is easily done using~\eref{gf}. Using also the explicit value of $\zeta_0$ \eref{zeta0}, 
the logarithmic negativity of two adjacent intervals, in the thermodynamic limit, is 
\begin{equation}
\label{neg-th}
{\cal E}_{A_1:A_2}=\ln\frac{G^2(e^{-\epsilon})}{G(e^{-2\epsilon})}. 
\end{equation}
%
%
\begin{figure}[t]
\begin{center}
\includegraphics*[width=0.65\linewidth]{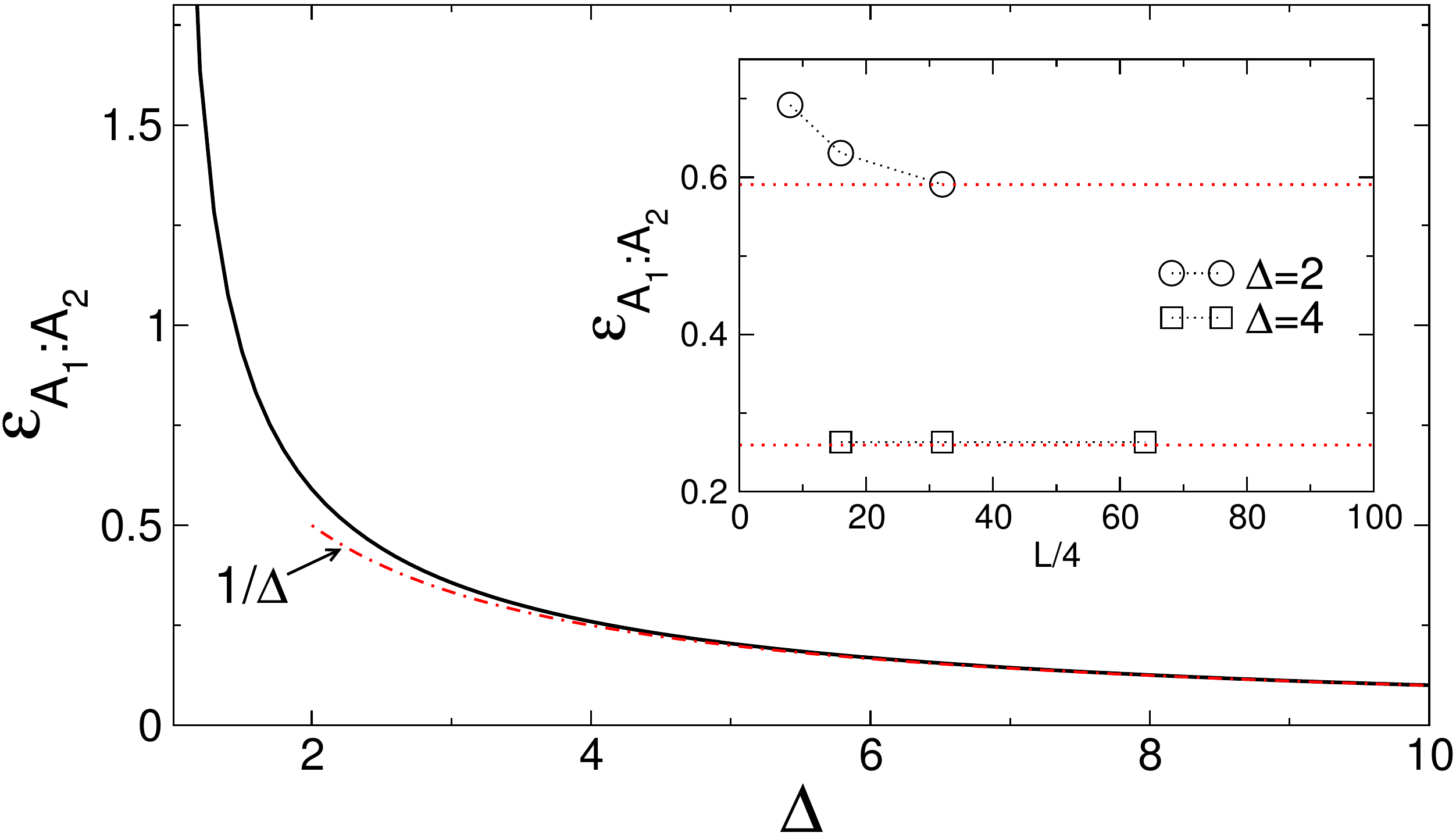}
\end{center}
\caption{Exact results for the logarithmic negativity $\cal E$ of two adjacent intervals in the XXZ spin 
 chain in the thermodynamic limit plotted as a function of the anisotropy $\Delta$. 
 The dash-dotted line is first order expansion for $\Delta\to\infty$. Inset: 
 Comparison with numerics for $\Delta=2$ and $\Delta=4$. The symbols are TEBD results 
 plotted as a function of the intervals length $L/4$ with $L$ the chain length. The dotted 
 lines are the theoretical predictions in the thermodynamic limit. 
}
\label{fig3}
\end{figure}
%
Via the relation $e^{-\epsilon}=\Delta-\sqrt{\Delta^2-1}$, this gives  the behaviour of ${\cal E}_{A_1:A_2}$ as a function of the 
anisotropy $\Delta$ as  shown in Figure~\ref{fig3}. In the limit $\Delta\to\infty$, 
${\cal E}_{A_1:A_2}$ vanishes, as a consequence of the vanishing of the bulk correlation 
length. The expansion for large $\Delta$ gives ${\cal E}_{A_1:A_2}=1/\Delta$, which 
is reported as dash-dotted line in Figure~\ref{fig3}, and it accurately describes the behaviour 
of ${\cal E}_{A_1:A_2}$  for $\Delta\gtrsim 4$. 
The inset in Figure~\ref{fig3} shows the comparison between finite-size TEBD results for ${\cal E}_{A_1:A_2}$ 
plotted as a function of the chain size $L$, and the thermodynamic limit result~\eref{neg-th}. 
While finite-size corrections are present for the smaller chains, it is clear that as $L\to\infty$ the data approach~\eref{neg-th}. 

We now discuss the moments of the partial transpose $\textrm{Tr}(\rho_{A_1\cup A_2}^{T_2}
)^n$. Using~\eref{gf}, their value in the thermodynamic limit is 
\begin{equation}
\label{traces}
\fl\quad\ln\textrm{Tr}(\rho_{A_1\cup A_2}^{T_2})^n=
\left\{\begin{array}{ll}
(1-n)\ln 2-3n\ln G(e^{-2\epsilon})+2\ln G(e^{-n\epsilon})G(e^{-2n\epsilon}) & n\,\textrm{even},\\
(1-n)\ln 2-3n\ln G(e^{-2\epsilon}) +3\ln G(e^{-2n\epsilon})      & n\,\textrm{odd}.
\end{array}\right.
\end{equation}
Note the dependence on the parity of $n$, alike systems described by conformal field theory~\cite{calabrese-2012}. 
By taking the analytic continuation $n\to 1$ from the result for $n$ even, the 
logarithmic negativity~\eref{neg-th} is recovered.  
On the other hand the analytic continuation for $n$ odd gives the normalisation 
$\textrm{Tr}\rho_{A_1\cup A_2}^{T_2}=1$, as it should.

\section{Critical behaviour close to $\Delta=1$ and unusual scaling corrections}
\label{unusual-sec}

We now discuss the critical behaviour of the moments of the partial transpose and of the 
logarithmic negativity in the vicinity of the critical point at $\Delta=1$, and  we also study the scaling corrections.
Unusual scaling corrections are known to affect entanglement-related quantities~\cite{ccen-2010,ccp-10,cardy-2010}. 
In contrast with standard scaling corrections, their exponents depend on the replica index $n$. 
Moreover, while usual scaling corrections arise due to bulk irrelevant operators, the unusual ones are due to the local insertion of a relevant operator  near the conical singularities \cite{cardy-2010,ot-2015} introduced to calculate entanglement in the replica trick \cite{cc-04, cfw-1994, hlw-1994}.
They are generically expected to appear also in gapped systems in the vicinity of a quantum critical point, as shown for the 
entanglement entropies \cite{ccp-10} and as we are going to show here for the moments of the partial transpose.

For the gapped XXZ chain the relation between the correlation length and the anisotropy $\Delta$ is 
known exactly. In the limit $\Delta\to 1$ (equivalently $\epsilon\to 0$), i.e., 
close to the conformal quantum critical point, one has 
\begin{equation}
\label{corr-l}
\ln \xi=\frac{\pi^2}{2\epsilon}+{\mathcal O}(\epsilon^0). 
\end{equation}
The strategy to derive the critical behaviour of the moments of the partial transpose is to 
consider the limit $\epsilon\to 0$, expanding~\eref{traces} in powers of $\xi$. 
Similar to the moments of the reduced density matrix~\cite{ccp-10}, the relations in~\eref{traces} 
are not suitable for taking the limit $\epsilon\to 0$. 
It is convenient to rewrite~\eref{traces} using the Poisson summation formula
\begin{equation}
\label{poisson}
\sum\limits_{j=-\infty}^\infty g(|\epsilon j|)=\frac{2}{\epsilon}\sum
\limits_{k=-\infty}^\infty\hat g\Big(\frac{2\pi k}{\epsilon}\Big), 
\end{equation}
where $\hat g(y)$ is the cosine transform of $g(x)$, defined as 
\begin{equation}
\hat g(y)\equiv \int_{0}^\infty g(x)\cos(yx)dx.
\end{equation}
Thus, defining  
\begin{equation}
h_n(x)\equiv\ln(1+e^{-n x}),\quad\textrm{with}\quad 
\hat h_n(y)=\frac{n}{2y^2}-\frac{\pi}{2y}\textrm{csch}\frac{\pi y}{n},
\end{equation}
one has 
\begin{equation}
\label{step}
\fl\quad
\ln G(z^n)=\frac{1}{2}\sum\limits_{j=-\infty}^\infty 
h_{n}(|j\epsilon|)-\frac{\ln 2}{2} =\sum\limits_{k=-\infty}^\infty
\Big[\frac{n\epsilon}{8\pi^2k^2}-\frac{1}{4k}\textrm{csch}\frac{2\pi^2 k}
{n\epsilon}\Big]-\frac{\ln 2}{2}, 
\end{equation}
where in the last step we used~\eref{poisson}. Plugging~\eref{step} in~\eref{traces}, 
one obtains for odd $n_o$
\begin{eqnarray}
\label{one}
\fl
\ln\textrm{Tr}(\rho_{A}^{T_2})^{n_o}=
-\frac{\pi^2}{8\epsilon}\Big(n_o-\frac{1}{n_o}\Big) +\frac{1}{2}(n_o-1)\ln2
+3\sum\limits_{k=1}^\infty\Big[\frac{n_o}{2k}
\textrm{csch}\frac{\pi^2k}{\epsilon}-\frac{1}{2k}\textrm{csch}
\frac{\pi^2k}{n_o\epsilon}\Big],
\end{eqnarray}
while for $n_e$ even one has  
\begin{eqnarray}
\label{two}
\fl
\ln\textrm{Tr}(\rho_{A}^{T_2})^{n_e}=
-&\frac{\pi^2}{8\epsilon}\Big(n_e-\frac{2}{n_e}\Big)+\frac12(n_e-2)\ln2\\\nonumber
&+\sum\limits_{k=1}^\infty\Big[
\frac{3n_e}{2k}\textrm{csch}\frac{\pi^2k}{\epsilon}-\frac{1}{k}
\textrm{csch}\frac{\pi^2k}{n_e\epsilon}-\frac{1}{k}\textrm{csch}
\frac{2\pi^2k}{n_e\epsilon}\Big].
\end{eqnarray}
In~\eref{one} and \eref{two} the terms $\propto1/\epsilon$ correspond to $k=0$. 
From~\eref{one} and \eref{two}, using~\eref{corr-l}, one obtains that 
$\ln\textrm{Tr}(\rho_A^{T_2})^n$ diverges in the limit $\epsilon\to 0$ as 
\begin{equation}
\ln\textrm{Tr}(\rho_A^{T_2})^n\simeq\left\{
\begin{array}{cl}
-\frac{1}{4}\big(n_o-\frac{1}{n_o}\big)\ln\xi +O(\xi^0) & n_o\;\textrm{odd},\\
-\frac{1}{4}\big(n_e-\frac{2}{n_e}\big)\ln\xi +O(\xi^0) & n_e\;\textrm{even}.
\end{array}
\right.
\end{equation}
Notice that these results are equivalent to the CFT scaling \cite{calabrese-2012,cct-neg-long} after the replacing 
$\ell_{1,2}\to\xi$, as one would naively expect from a simple scaling reasoning. 
Taking the limit $n_{e/o}\to\infty$ one gets the scaling of the largest eigenvalue of $\textrm{Tr}(\rho_A^{T_2})^n$
\begin{equation}
\ln\lambda_0=-\frac{\pi^2}{8\epsilon}+\frac{\ln2}2 +O(\epsilon), 
\end{equation}
which agrees with the expansion of \eref{lambda0} close to $\epsilon=0$.
The scaling of the logarithmic negativity is instead 
\begin{equation}
{\cal E}_{A_1:A_2}=\lim_{n_e\to1} \ln\textrm{Tr}(\rho_{A}^{T_2})^{n_e}=
\frac{\pi^2}{8\epsilon}-\frac{\ln2}2 +O(\epsilon)\,,
\end{equation}
which agrees with the expansions of \eref{neg-th}.

Finally, the sums in~\eref{one} and \eref{two} give rise to scaling corrections in the limit 
$\epsilon\to0$. Using the asymptotic behaviour $\textrm{csch}(x)\propto e^{-x}$ for $x\to+\infty$,  
one obtains corrections of the form $\xi^{-2k}$ from the first term in the square brackets 
in~\eref{one} and \eref{two}. The last two terms in~\eref{one} and \eref{two} give corrections of the 
form $\xi^{2k/n}$ and $\xi^{4k/n}$ which are the ``expected'' unusual corrections, i.e., 
they have the same exponents appearing in the corrections for the moments of the reduced density matrix \cite{ccp-10}.

\section{Negativity spectrum distribution}
\label{neg-dist}

In this section we investigate the distribution of the negativity spectrum which is defined as
\begin{equation}
\label{P}
P(\lambda)\equiv\sum\limits_{i}\delta(\lambda-\lambda_i), 
\end{equation}
as sum over the eigenvalues $\lambda_i$ of the partial transpose.
So far, this has been investigated  only for conformal field theories~\cite{ruggiero-2016}.  
Using the explicit expressions of the eigenvalues of the partial transpose and their degeneracy, 
in the XXZ spin chain $P(\lambda)$ can be written as
\begin{equation}
\label{P1}
P(\lambda)=\sum_{k=0}^\infty\delta(\lambda-\lambda_0 e^{-k \epsilon})q_+(k)+\sum_{k=0}^\infty\delta(\lambda-\lambda_0 e^{-k \epsilon})q_-(k) ,
\end{equation}
with $\lambda_0$ being the largest eigenvalue of the partial transpose \eref{lambda0}. 
In the limit of large $k$ (i.e. for small $\lambda$ of the partial transpose)~\eref{P} approaches a continuous distribution 
which can be determined from the asymptotic behaviour of the degeneracy $q_\pm(k)$ for large $k$. 
In ~\ref{mein-app} we use the Meinardus theorem to  obtain the large $k$ behaviours
\begin{equation}
\label{qpm}
q_\pm(k)\simeq\left\{
\begin{array}{cc}
\frac{k^{-\frac{3}{4}}}{8\sqrt{2}}(e^{\pi\sqrt{k}}\pm 2^{\frac{5}{4}}e^{\frac{\pi}{
\sqrt{2}}\sqrt{k}}) & \textrm{even}\, k\\\\\frac{k^{-\frac{3}{4}}}{8\sqrt{2}}e^{\pi\sqrt{k}} 
& \textrm{odd}\, k.
\end{array}
\right.
\end{equation}
Consequently one can  write
\begin{equation}\fl
P(\lambda_k) \frac{d\lambda_k}{dk}=  
\theta(\lambda_k) \frac{k^{-\frac{3}{4}}}{8\sqrt{2}}(e^{\pi\sqrt{k}}+ 2^{\frac{1}{4}}e^{\frac{\pi}{\sqrt{2}}\sqrt{k}})
+\theta(-\lambda_k) \frac{k^{-\frac{3}{4}}}{8\sqrt{2}}(e^{\pi\sqrt{k}}- 2^{\frac{1}{4}}e^{\frac{\pi}{\sqrt{2}}\sqrt{k}}),
\label{Pk}
\end{equation}
where $\lambda_k=\lambda_0 e^{-k\epsilon}$ is the $k$-th eigenvalue. 
The approximation of the sum in~\eref{P} with an integral  requires that $\ln(\lambda_0/|\lambda_k|)\gg\epsilon$. 
Since $\epsilon=\textrm{arccosh}\Delta$, this implies that 
the larger $\Delta$ is, the smaller $|\lambda_k|$ has to be for~\eref{Pk} to be valid. 

To make contact with Ref.~\cite{ruggiero-2016}, it is worth considering 
the  number distribution function $n(\lambda)$, which  for $\lambda>0$ ($\lambda<0$) counts the number of eigenvalues 
of the partial transpose larger (smaller) than $\lambda$. 
It is useful to treat the cases $\lambda>0$ and $\lambda<0$ separately, defining $n_{\pm} (|\lambda|)$ as 
\begin{equation}
\label{npm}
n_\pm(|\lambda|)=\sum_{k=0}^\infty\theta(|\lambda| - \lambda_0e^{-k\epsilon}) q_{\pm}(k)=
\sum\limits_{k=0}^\infty\theta(\ln(\lambda_0/|\lambda|)-k\epsilon)q_\pm(k).
\end{equation}
The asymptotic behaviour of $n_\pm(|\lambda|)$ is straightforwardly obtained from that of $P(\lambda)$, 
and it is conveniently written in the suggestive form
\begin{equation}
\label{ngap}
n_\pm(\lambda)=\frac{1}{\sqrt{2\eta}\pi}\Big(\frac{1}{4}e^{\pi\eta}
\pm\frac{1}{2^\frac{5}{4}}e^{\frac{\pi}{\sqrt{2}}\eta}\Big),\quad\textrm{with}\;\;
\eta\equiv\sqrt{\frac{\ln(\lambda_0/|\lambda|)}{\epsilon}},
\end{equation}
where all the dependence on $\lambda$ is encoded in the universal scaling variable $\eta$, 
without any explicit dependence on the parameters of the model. 
This is very similar to what found in conformal field theories \cite{ruggiero-2016}, 
but a more quantitative comparison between the two results requires a careful analysis (with consequences 
also on the entanglement spectrum) which will be discussed elsewhere \cite{act-prep}.

%
\begin{figure}[t]
\begin{center}
\includegraphics*[width=0.75\linewidth]{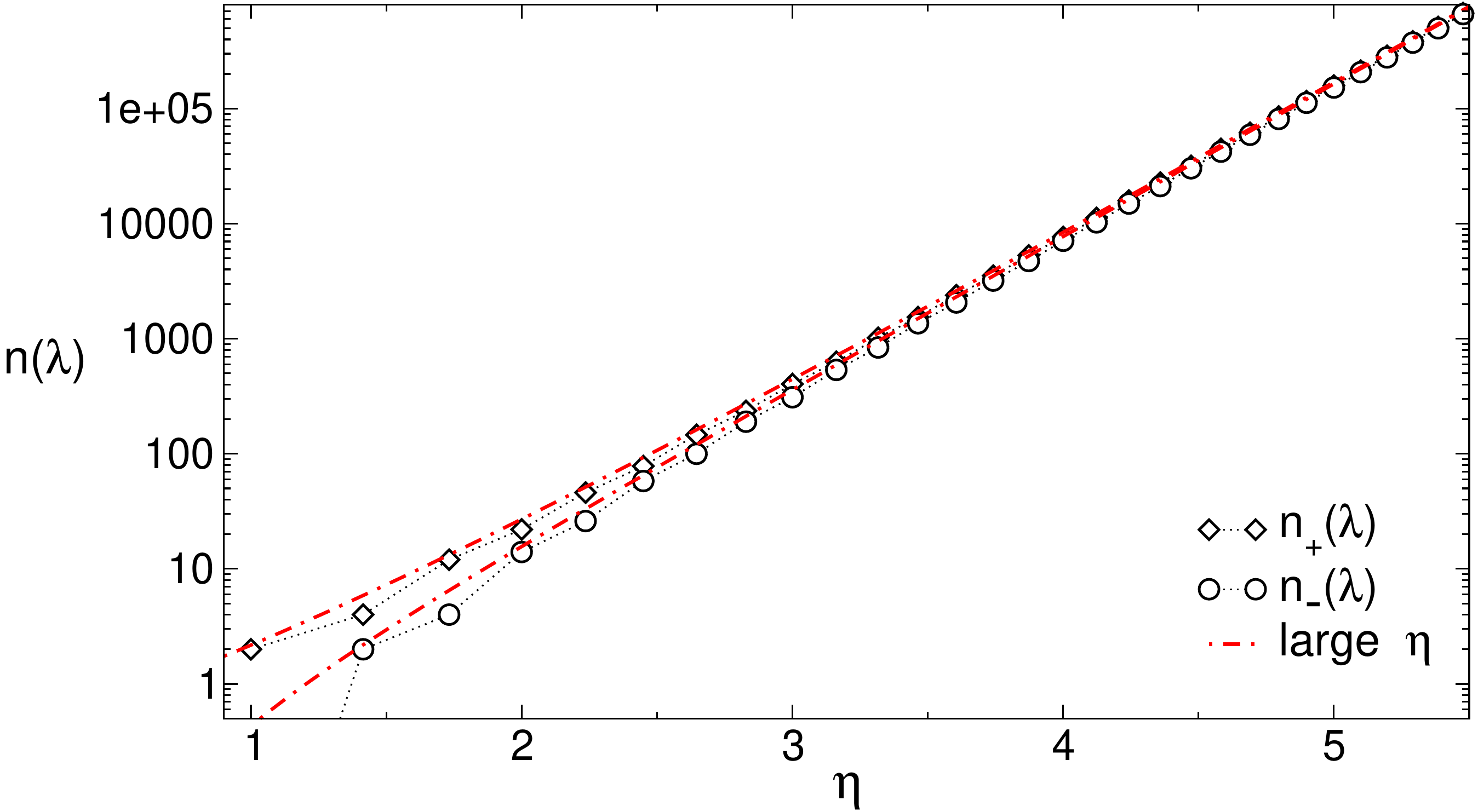}
\end{center}
\caption{Negativity spectrum of two adjacent intervals in the XXZ chain: number distribution $n(\lambda)$ plotted as a function 
 of $\eta=((\ln(\lambda_0/|\lambda|)/ \epsilon)^\frac{1}{2}$, which does not  dependent on $\Delta$. 
 The symbols correspond to the exact result in the thermodynamic limit given by \eref{npm}.
 Diamonds (circles) correspond to positive (negative) eigenvalues. 
 The continuous lines are the asymptotic behaviours for large $\eta$ \eref{ngap} which describe well the data for $\eta\gtrsim2$. 
}
\label{fig2}
\end{figure}
%

%
In the limit $\eta\to\infty$ both $n_+(\lambda)$ and $n_-(\lambda)$ exhibit the same asymptotic behaviour as
\begin{equation}
\label{nasy}
n_\pm(\lambda)\propto\frac{1}{4\pi\sqrt{2\eta}}\exp({\pi\eta}),
\end{equation}
which means that the number distribution does not depend on the sign of the eigenvalues of the 
partial transpose. Similar to~\eref{nasy}, for the negativity spectrum of conformal 
field theories~\cite{ruggiero-2016a} the distribution of the small eigenvalues 
of the partial transpose is exponential, and it does not depend on the sign of the 
eigenvalues. 
The same asymptotic behaviour for $|\lambda|\to0$ is already visible for 
$\eta\equiv(\ln(\lambda_0/|\lambda|)/\epsilon)^{\frac{1}{2}}\approx 4$, see Figure~\ref{fig2}.

We now provide numerical evidence supporting the analytical result for the number 
distribution $n_\pm(\lambda)$ in the gapped XXZ chain. 
It is not instructive to compare \eref{ngap} with the TEBD results since we already showed that 
these agree well with the CTM prediction in the thermodynamic limit and so we 
prefer to contrast the exact CTM result to the asymptotic prediction. 
This comparison is illustrated in Figure~\ref{fig2}, plotting $n_+(\lambda)$ and $n_-(\lambda)$ versus 
the scaling variable $\eta=(\ln(\lambda_0/|\lambda|)/\epsilon)^\frac{1}{2}$, 
which is a plot independent from $\Delta$. 
The dash-dotted lines are the analytical results~\eref{ngap}, and are in excellent 
agreement with the numerics already for $\eta\gtrsim 2$.

\section{Conclusions}
\label{concl}

We investigated the negativity spectrum of two adjacent intervals in generic gapped 
one-dimensional systems. We demonstrated that in the limit of large intervals the 
negativity spectrum can be reconstructed completely from the entanglement spectrum of 
the bipartite system. Our main result is formula~\eref{neg_spec}. The result 
relies upon two main physical ingredients. First, the contributions to the negativity 
spectrum originating from different boundaries (namely, the boundary between the 
two subsystems and the subsystems and the rest) are factorized. Second, due to 
the finite correlation length, around the boundary between the two intervals the 
system is described effectively by a pure state. 

As an application of our result we discussed in detail the negativity spectrum of the 
spin-$\frac{1}{2}$ Heisenberg XXZ chain in the gapped phase for $\Delta>1$. 
We considered separately the negative and positive eigenvalues of the partial 
transpose. In both cases the negativity spectrum levels are equally spaced, 
similar to the entanglement spectrum. The level spacing, however, is half that of 
the entanglement spectrum. Negativity spectrum levels are highly degenerate. 
The degeneracies can be characterised analytically in terms of the partition of the 
integers. This allowed us to provide exact results for the logarithmic negativity, 
the moments of the partial transpose, and the associated unusual scaling corrections. 
Finally, we derived analytically the distribution of the negativity spectrum levels. 
We found that the  number distribution is independent of $\Delta$. Moreover, 
for the small eigenvalues of the partial transpose it does not depend on the eigenvalue 
sign, as for the negativity spectrum of conformal field theories~\cite{ruggiero-2016a}. 
We provided accurate numerical checks of our results using TEBD simulations. 

We finish by discussing some directions for future work. First, we considered 
the negativity spectrum of the gapped XXZ chain, which is an integrable model
(but our results are valid for an arbitrary model solvable by corner transfer matrix as e.g.
those in Refs. \citenum{w-06,eer-10,eefr-12,df-13,br-16}). 
It would be interesting to generalise our results for the distribution 
of the entanglement spectrum levels (for instance~\eref{ngap}) to non-integrable systems. 
Moreover, the negativity spectrum distribution could have applications also to disordered models exhibiting many body localisation, 
generalising the results for the entanglement spectrum~\cite{serbyn-2016}. 
Finally, it would also interesting to clarify the signatures of topological order in the negativity spectrum. 

\section{Acknowledgments}
V.~A.~acknowledges support from the European Union's Horizon 2020 research and innovation programme under the
Marie Sklodowoska-Curie grant agreement No 702612 $OEMBS$.

\appendix
\section{Meinardus theorem}
\label{mein-app}

Here we review the main result of the Meinardus theorem (see Ref.~\citenum{andrews-book} 
for more details). The power of Meinardus theorem is that it allows to obtain the asymptotic 
behaviour of a large class of restricted integer partition $r(k)$ for large $k$. Specifically, let 
us consider a generic $r(k)$ that is defined via the generating function 
$f(q)$ as 
\begin{equation}
\label{fq}
f(q)\equiv 1+\sum_{k=1}^\infty r(k)q^k=\prod_{k=1}^\infty(1-q^k)^{-\alpha_k}, 
\end{equation}
where $|q|<1$ and $\alpha_k$ are non-negative numbers. Different types of restricted partitions 
can be obtained by choosing different sequences $\alpha_k$. It is convenient for the following 
to define an auxiliary Dirichlet series $D(s)$ as 
\begin{equation}
\label{dir}
D(s)\equiv\sum_{n=1}^\infty \frac{\alpha_n}{n^s}. 
\end{equation}
The Meinardus theorem states that the leading behaviour of $r(k)$ for large $k$ is given as 
\begin{equation}
\label{mein}
r(k)\propto C n^\kappa\exp\Big\{n^{\alpha/(\alpha+1)}\Big(1+\frac{1}{\alpha}
\Big)[A\Gamma(\alpha+1)\zeta(\alpha+1)]^{1/(\alpha+1)}\Big\}.
\end{equation}
Here
\begin{equation}
C\equiv e^{D'(0)}[2\pi(1+\alpha)]^{-\frac{1}{2}}[A\Gamma(\alpha+1)\zeta(\alpha+
1)]^{(1-2D(0))/(2+2\alpha)}, 
\end{equation}
$\kappa$ is given as 
\begin{equation}
\kappa=\frac{D(0)-1-\alpha/2}{1+\alpha}, 
\end{equation}
and $A$ is the residue of the first order pole of $D(s)$ (cf.~\eref{dir}) at $s=\alpha$.

We now proceed to apply the Meinardus theorem to the large $k$ behaviour of $q_{\pm}(k)$ of our interest. 
To proceed, it is convenient to first derive the asymptotic behaviour of the combinations 
$q_+\pm q_-$. From~\eref{gf}, one has that the generating function of $(q_++q_-)/2$ is 
$G^2(x^2)G^2(x)$, 
\begin{equation}
G^2(x^2)G^2(x)=\prod_{k=1}^\infty (1+x^{2k})^2(1+x^k)^2=\prod\limits_{k=1}^\infty 
(1-x^k)^{\alpha'_k}, 
\end{equation}
where $\alpha'_k=2$ for $k$ not divisible by four, and zero otherwise.  
Using~\eref{dir}, one has 
\begin{equation}
D(s)=2^{1-2s}(2^{2s}-1)\zeta(s). 
\end{equation}
$D(s)$ has a pole of order one at $s=1$ and residue $A=3/2$, implying $\alpha=1$ 
in~\eref{mein}. Using the Meinardus theorem, one obtains  
\begin{equation}
\label{p}
q_++q_-\propto \frac{1}{4\sqrt{2}}k^{-\frac{3}{4}}e^{\pi\sqrt{k}},\quad k\to\infty. 
\end{equation}
On the other hand, one has that the generating function for $(q_+-q_-)/2$ is 
\begin{equation}
G^3(x^2)=\prod\limits_{k=1}^\infty(1-x^{k})^{\alpha''_k}, 
\end{equation}
where $\alpha''_k=3$ if $k \bmod 4 = 2$, and zero otherwise. 
It is straightforward to check that in this case 
\begin{equation}
D(s)=\frac{3}{2^{2s}}(2^s-1)\zeta(s). 
\end{equation}
This implies that $\alpha=1$ and $A=3/4$ in~\eref{mein}. 
The Meinardus theorem now yields 
\begin{equation}
\label{m}
q_+-q_-\propto 
\left\{\begin{array}{cc}
2^{-\frac{5}{4}}k^{-\frac{3}{4}}e^{\frac{\pi}{\sqrt{2}}\sqrt{k}} &  
k\,\textrm{even}\\
0 & k\,\textrm{odd}.
\end{array}\right. 
\end{equation}
Using~\eref{p} and~\eref{m}, one obtains for large $k$
\begin{equation}
\label{qpm1}
q_\pm(k)\simeq\left\{
\begin{array}{cc}
\frac{k^{-\frac{3}{4}}}{8\sqrt{2}}(e^{\pi\sqrt{k}}\pm 2^{\frac{5}{4}}e^{\frac{\pi}{
\sqrt{2}}\sqrt{k}}) & \textrm{even}\, k\\\\\frac{k^{-\frac{3}{4}}}{8\sqrt{2}}e^{\pi\sqrt{k}} 
& \textrm{odd}\, k,
\end{array}
\right.
\end{equation}
which is the result reported in the main text (cf.~\eref{qpm}).


\end{document}